\documentclass{article}
\begin{document}
\title{DYNAMICS AND COOLING OF MODULATED BUNCHES IN BEND}
\author{  Rafael Tumanyan, \\
Yerevan Physics Institute, Armenia }
\maketitle
\begin{abstract}

The influence of the space charge fields 
on the dynamics of the particles of the bunch, which is moved in a bending 
magnet is calculated. It is shown, that such influence of the Coulomb force on 
the energy of the particles of a modulated beam causes in the decreasing of 
the bunch energy spread during the compression due to energy deviation 
along the compressed bunch, i.e. bunch cooling. The necessary conditions 
of the maximal cooling are found. The attainable ordering of the bunch is investigated.

Prepared for EPAC04
\end{abstract}

\section{  INTRODUCTION}
The cooling of charged particles beams is very impotant issues. The laser cooling 
\cite{raf} have many advantages in comparison to other cooling methods and very fast. 
But sometimes it is necessary to use more usual cooling.  
The effects of short electron bunch self fields  on the bunch emittance and 
particles motion become important last years due to achievements in reception 
of short (shorter than 0.1 mm) bunches by using magnetic compression \cite{2,7}. 
This paper is devoted to dynamics of the bunch under the influence of the other 
cooperative charge field effects - Coulomb and centripetal force and collective 
focusing forces. All these forces grow, when the bunch length decreases. We 
have find that in this case the cooling of the bunch is possible.

\section {GENERAL CONSIDERATION}
The Coulomb force in the bunch with linear charge density 
$\lambda (s)$
, which is moved along the z- axe with velocity 
$V=c\beta $
, is determined by Poisson equation 
$$
\Delta \varphi  = 4\pi \lambda (s)(1 - \beta )
$$
In the approach $\beta=const$ this equation for electrical potential has the solution
\begin{equation}
\varphi=\frac{4\pi}{2\gamma^2}\int ds\int ds'\lambda(s')
\label{elpot}
\end{equation}
and we have for Coulomb force 
\begin{equation}
F_C=-\frac{d\varphi}{ds}=\frac{4\pi}{2\gamma^2}\int ds'\lambda(s')
\label{Cforce}
\end{equation}
For uniform bunches (
$\lambda (s) = const$
) with the number of particles N and length 
$\sigma $
, this equation has the solution
\begin{equation}
\varphi _c  = 4\pi eN/\sigma 2\gamma ^2 
\label{coul}
\end{equation}
This solution with the solution of the CSR \cite{7,6} in the bend (\ref{enloss}) and (\ref{unifII}) 
give us the term, which is proportional to $\gamma^{-2}$ and is not considered in \cite{7,6}.
But this one is important in case of lower energy ($\gamma <100$) of the bunch and 
describes the influence of  the bunch compressing on the energy deviation 
of the  bunch. We find in this paper the conditions of bunch cooling during the compression 
of modulated bunches in the magnetic fields such as alfa-magnet or chicane. 
For this goal we  derive now particle dynamics equations for a relativistic 
(
$\beta  = {\raise0.7ex\hbox{$V$} \!\mathord{\left/
 {\vphantom {V c}}\right.\kern-\nulldelimiterspace}
\!\lower0.7ex\hbox{$c$}} \approx 1$
) bunch that follows curved trajectory with radius R. If a particle at the tail of 
the bunch radiates the electromagnetic fields, then the radiation propagates 
along the chord and ``catches'' another particle after overtaking distance \cite{6} of
$L_0  = \left| {AB} \right| = \theta R = 2(3sR^2 )^{{\raise0.7ex\hbox{$1$} \!\mathord{\left/
 {\vphantom {1 3}}\right.\kern-\nulldelimiterspace}
\!\lower0.7ex\hbox{$3$}}}$
(s is the distance between particles). Under the influence of the radiation forces 
the particle motion nearby an equilibrium trajectory (for a given energy and 
given momentum) can be described by the following equations: 
\begin{equation}
x'' + (K^2  - n)x = K\frac{{\Delta \varepsilon }}
{\varepsilon } + \frac{{F_x }}
{\varepsilon }
\label{baseq}
\end{equation}
$$
y'' + ny = {\raise0.7ex\hbox{${F_y }$} \!\mathord{\left/
 {\vphantom {{F_y } \varepsilon }}\right.\kern-\nulldelimiterspace}
\!\lower0.7ex\hbox{$\varepsilon $}},\varepsilon ' = e{\raise0.7ex\hbox{${\vec E\vec V}$} \!\mathord{\left/
 {\vphantom {{\vec E\vec V} {v_x }}}\right.\kern-\nulldelimiterspace}
\!\lower0.7ex\hbox{${v_x }$}},t' = {\raise0.7ex\hbox{${(1 + Kx)}$} \!\mathord{\left/
 {\vphantom {{(1 + Kx)} {v_x }}}\right.\kern-\nulldelimiterspace}
\!\lower0.7ex\hbox{${v_x }$}}
$$
where 
$\left( ' \right) = {\raise0.7ex\hbox{$d$} \!\mathord{\left/
 {\vphantom {d {dx}}}\right.\kern-\nulldelimiterspace}
\!\lower0.7ex\hbox{${dx}$}},K(z) = {\raise0.7ex\hbox{$1$} \!\mathord{\left/
 {\vphantom {1 R}}\right.\kern-\nulldelimiterspace}
\!\lower0.7ex\hbox{$R$}}$
 is the equilibrium orbit curvature, n(z) is the external focusing quadrupole 
field index. We neglected terms of  
$y'\varepsilon ',x'\varepsilon '$
 because no essential parametric damping is assumed. 
  The components of the Lorenz force 
$\vec F = e(\vec E + \vec \beta  \times \vec B)$
 can be calculated via the electromagnetic potential 
 (
$A_0 ,A_x ,A_y  = 0,A_z $
) as 
\begin{equation}
F_x  =  - \frac{{\partial V_0 }}
{{\partial x}} - e\frac{{dA_x }}
{{cdt}} + e\frac{{KA_x }}
{{1 + Kx}}
\label{silafx}
\end{equation}
\begin{equation}
F_y  =  - \frac{{\partial V_0 }}
{{\partial y}},e\vec E\vec \beta  = \frac{{\partial V_0 }}
{{\partial t}} - e\frac{{dA_0 }}
{{dt}}
\label{silafy}
\end{equation}
where 
$U_0  = e(A_0  - \beta A_x )$
is the interaction potential. To work out the perturbation of electron motion 
under the effect of self-radiative forces eqs. (\ref{silafx},\ref{silafy}) above we 
transform the eq. (\ref{baseq}) into equations for amplitudes 
$C_{x,y} $
of Floquet solutions  
$f_{x,y} $
 of the equation (\ref{baseq}) without right-hand part. In this notation the full 
 solution of (\ref{baseq}) is may be written in the following form
$x = \Psi \frac{{\Delta \varepsilon }}
{\varepsilon } + (C_x ^ *  f_x  + c.c.),y = C_y^ *  f_y  + c.c.$, 
here $\Psi$ is the solution of equation (\ref{baseq}) with right-hand part equal to K. 
Floquet functions 
$f_{x,y} $
 satisfy to normalization condition 
$f^{'} _{x,y} f_{x,y}^ {*}   - c.c. = 2i$.
Now it is not difficult to find the complex amplitudes (integrals of unperturbed motion) 
as functions of coordinates, velocities and energy: 
$$
2iC_x  = xf'_x  - x'f_x  + \eta \frac{{\Delta \varepsilon }}
{\varepsilon },2iC_y  = yf'_y  - y'f_y 
$$
where 
$\eta (z) = \int\limits_{ - \infty }^z {f_x Kdz} $. 
We can obtain time derivatives if we take into account equations of perturbed motion 
\begin{equation}
2iC'_y  =  - f_y ({{F_y } \mathord{\left/
 {\vphantom {{F_y } \varepsilon }} \right.
 \kern-\nulldelimiterspace} \varepsilon })
 \label{cy}
\end{equation}
\begin{equation}
2iC'_x  =  - f_x ({{F_x } \mathord{\left/
 {\vphantom {{F_x } \varepsilon }} \right.
 \kern-\nulldelimiterspace} \varepsilon }) + \eta {\raise0.7ex\hbox{${\varepsilon '}$} \!\mathord{\left/
 {\vphantom {{\varepsilon '} \varepsilon }}\right.\kern-\nulldelimiterspace}
\!\lower0.7ex\hbox{$\varepsilon $}}
\label{cx}
\end{equation}
The final equations for displaced amplitudes
$\hat C_y  = C_y $,
$\hat C_x  = C_x  + e\frac{{\eta A_0 }}
{{2i\varepsilon }} - \frac{{f_x A_x }}
{{2i\varepsilon }}$
and energy 
$\hat \varepsilon  = \varepsilon  + eA_0 $
 may be written in the following form
$$
2i\varepsilon \hat C'_x  = (\frac{\eta }
{c}\frac{\partial }
{{\partial t}} + f_x \frac{\partial }
{{\partial x}})U_0  - ef'_x A_x ,
$$
$$
2i\varepsilon \hat C'_y  = f_y \frac{{\partial U_0 }}
{{\partial y}};\hat \varepsilon '_{}  = \frac{1}
{c}\frac{\partial }
{{\partial t}}U_0 (1 + Kx).
$$
\section {TRANSVERSE FORCES OF CSR}

To calculate the effective transverse forces and perturbation of the amplitudes, 
we assumed, that the bunch length $\sigma$ satisfies conditions of a "thin" bunch and 
absence of beam pipe shielding \cite{6}. Using the following integrals for electromagnetic 
potentials 
$$
A_0  = \int {\frac{{d^3 \vec r_1 }}
{{c\tau }}} \rho (\vec r_1 ,l - \tau );\tau  = {{\left| {\vec r_1  - \vec r} \right|} \mathord{\left/
 {\vphantom {{\left| {\vec r_1  - \vec r} \right|} c}} \right.
 \kern-\nulldelimiterspace} c}
$$
$$
\vec A = \int {\frac{{d^3 \vec r_1 }}
{{c\tau }}\vec j(\vec r_1 ,l - \tau );\vec j(\vec r,l) = \vec \beta \rho } (\vec r,l)
$$
here $\rho$ is the charge density distribution of the bunch. The retarding 
distance $c\tau$ for a 
constant bending radius R can be presented in the form as 
$$
\left| {\vec r_1  - \vec r} \right| = \left| {\zeta  - \frac{{\zeta ^3 }}
{{24R^2 }} + \zeta ^{} \frac{{x + x_1 }}
{{2R}} + \frac{{(x - x_1 )^2  + (y - y_1 )^2 }}
{{2\zeta }}} \right|,\zeta  = z - z_1 
$$
 We derive in this paper the effective transverse forces with linear accuracy 
on x and y. We neglect here the integration over  
$\zeta  \leq 0$,
small ultra­relativistic terms 
$ \propto \gamma ^{ - 2} $
 and transverse dispersion of $\tau$ in the denominator of the integrand in retarding 
 potentials. In assumption that $\rho$ is even function and in ignoring of small terms 
 of second order after integrating one can find for interaction potential 	
$$
U_0  = U(s)(1 + Kx) - F_0 x + \frac{1}
{2}g(s)(3x^2  + y^2 ),
$$
where 
$$
F_0 (s) =  - \frac{{2Ne^2 }}
{R}\lambda (s),U(s) = \frac{{2Ne^2 }}
{{(3R^2 )^{{1 \mathord{\left/
 {\vphantom {1 3}} \right.
 \kern-\nulldelimiterspace} 3}} }}\int\limits_{}^{} {\frac{{ds_1 }}
{{s_1^{{1 \mathord{\left/
 {\vphantom {1 3}} \right.
 \kern-\nulldelimiterspace} 3}} }}\lambda (s - s_1 )} 
$$
and notations $\lambda$ for linear charge density and 
$$
g(s) = \frac{{Ne^2 }}
{{(3R^2 )^{{2 \mathord{\left/
 {\vphantom {2 3}} \right.
 \kern-\nulldelimiterspace} 3}} }}\frac{\partial }
{{\partial s}}\int\limits_0^\infty  {\frac{{ds_1 }}
{{s_1^{{2 \mathord{\left/
 {\vphantom {2 3}} \right.
 \kern-\nulldelimiterspace} 3}} }}\lambda (s - s_1 )} 
$$
are used.  It is clear that the contribution of the radial vector potential to displaced 
amplitudes C is about small terms of order 
$ \sim (R^2 \sigma _z )^{{1 \mathord{\left/
 {\vphantom {1 3}} \right.
 \kern-\nulldelimiterspace} 3}} /\beta _x $
, with respect to 
$F_0 (s)$
 and g(s).
Hence, we can neglect them in further consideration. In this approach we have 
$$
2i\varepsilon \hat C'_x  =  - \eta \frac{\partial }
{{\partial s}}U(s) - f_x F_0 (s) + 3f_x g(s)x
$$
$$
2i\varepsilon \hat C'_y  = f_y g(s)y,\hat \varepsilon ' =  - \frac{\partial }
{{\partial s}}U(s)
$$
Consequently, comparing with initial Eqs. (\ref{cy},\ref{cx}), one can see that: 
1)   ${{\partial U} \mathord{\left/
  {\vphantom {{\partial U} {\partial s}}} \right.
\kern-\nulldelimiterspace} {\partial s}}$
  is longitudinal energy loss gradient originally found in [6],
 2)
  $F_0 (s)$
  is  the effective centripetal radial force, 

 3) terms with g(s) describe focusing field distortions in both transverse planes. 
 
 All these forces cause emittance growth. 

\section {PARTICULAR DISTRIBUTIONS OF BUNCH DENSITY}

For a bunch with Gaussian linear charge density distribution 
$\lambda (s) = ({1 \mathord{\left/
 {\vphantom {1 {\sqrt {2\pi } \sigma _s }}} \right.
 \kern-\nulldelimiterspace} {\sqrt {2\pi } \sigma _s }})e^{ - {{s^2 } \mathord{\left/
 {\vphantom {{s^2 } {2\sigma _s ^2 }}} \right.
 \kern-\nulldelimiterspace} {2\sigma _s ^2 }}} $
, the energy loss gradient along the bunch equals to 
\begin{equation}
\hat \varepsilon ' = \frac{{d\varepsilon }}
{{cdt}} =  - \frac{{2Ne^2 }}
{{\sqrt {2\pi } (3R^2 \sigma _s^4 )^{{1 \mathord{\left/
 {\vphantom {1 3}} \right.
 \kern-\nulldelimiterspace} 3}} }}I_0 ({s \mathord{\left/
 {\vphantom {s {\sigma _s }}} \right.
 \kern-\nulldelimiterspace} {\sigma _s }})
 \label{enloss}
\end{equation}
where the function 
$I_0 $  equals to
$$
I_0 (s) = \frac{\partial }
{{\partial s}}\int\limits_0^\infty  {\frac{{ds_1 }}
{{s_1^{1/3} }}} \lambda (s - s_1 ) = \frac{\partial }
{{\partial s}}e^{ - s^2 /2} H_{ - 2/3} (s)
$$

and 
$H_n (s)$
 is the Hermite function and may be expressed through the hypergeometric function 
$_1 F_1 $.
As it is qualitatively understood, the bunch head particles get some excess 
of energy, while the tail and the center part mostly loses the energy. Transverse 
forces within the Gaussian bunch are given by formulae: 
$$
F_x (s) =  - \frac{{2Ne^2 }}
{R}\lambda (s) - x\frac{{3Ne^2 }}
{{\sqrt {2\pi } (9R^4 \sigma _s^5 )^{{1 \mathord{\left/
 {\vphantom {1 3}} \right.
 \kern-\nulldelimiterspace} 3}} }}I_1 ({s \mathord{\left/
 {\vphantom {s {\sigma _s }}} \right.
 \kern-\nulldelimiterspace} {\sigma _s }})
$$
$$
F_y (s) =  - y\frac{{Ne^2 }}
{{\sqrt {2\pi } (9R^4 \sigma _s^5 )^{{1 \mathord{\left/
 {\vphantom {1 3}} \right.
 \kern-\nulldelimiterspace} 3}} }}I_1 ({s \mathord{\left/
 {\vphantom {s {\sigma _s }}} \right.
 \kern-\nulldelimiterspace} {\sigma _s }})
$$
where 
$$
I_1 (s) = \frac{\partial }
{{\partial s}}\int\limits_0^\infty  {\frac{{ds_1 }}
{{s_1^{2/3} }}} \lambda (s - s_1 ) = \frac{\partial }
{{\partial s}}e^{ - s^2 /2} H_{ - 1/3} (s)
$$


One can see that particles at the head of the bunch are defocused by overtaking 
radiation fields, while other particles are focused. Let us note that effective 
radial force 
${\raise0.7ex\hbox{${\partial U_0 }$} \!\mathord{\left/
 {\vphantom {{\partial U_0 } {\partial x}}}\right.\kern-\nulldelimiterspace}
\!\lower0.7ex\hbox{${\partial x}$}}$
 is essentially different from the initial force 
$F_x $
 in (2) -- the difference is the term  
$ \propto KA_z $
. In fact, the later term dominates in 
$F_x $
 and it is the centrifugal  force defined by Talman \cite{3} (it looks like 
$({{Ne^2 } \mathord{\left/
 {\vphantom {{Ne^2 } R}} \right.
 \kern-\nulldelimiterspace} R})\lambda (s)\ln \frac{{(R\sigma _s ^2 )^{{2 \mathord{\left/
 {\vphantom {2 3}} \right.
 \kern-\nulldelimiterspace} 3}} }}
{{\sigma _ \bot ^2 }}$
the effect of this later force on the bunch particles is cancelled by effect of 
the particle energies deviation under the influence of the transverse electric 
field, and therefore does not lead to the emittance growth. Similar cancellation 
effect in the particular case of a coasting beam was found in \cite{4}. Now we can 
conclude that it is valid for any relativistic bunch. Fig. shows the plot of $I_0$, which is 
the higher curve, and of $I_1$. It is seen, that the resulting force is not centered 
relative to bunch center.  
 This results are applicable to uniform bunches also, but after substitution
\begin{equation}
 I_0 (s) = (\frac{\sigma }{2} - s)^{ - 1/3} ;  
 I_1 (s) = (\frac{\sigma }{2} - s)^{ - 2/3} 
\label{unifII}
\end{equation}
 It is not difficult to see, that this formulas are diverge at the center of the bunch 
$s = \sigma /2$.
\section{ BUNCH COOLING IN THE BEND ACCOUNTING FOR COULOMB FORCES}
The solutions (\ref{elpot},\ref{coul}) with the solution of the CSR in the bend 
(\ref{enloss}) and (\ref{unifII}) 
give us the terms, which are proportional to $\gamma^{-2}$ and is not considered in \cite{7,6}.
But this one is important in case of lower energy ($\gamma <100$) of the bunch and 
describes the influence of  the bunch compressing on the energy deviation 
of the bunch. 
The necessary condition of bunch cooling by compression at way by total lenght 
$L$ by using of the bending magnet with lenght $L_m$can be found by 
comparing of maximal energy deviation at the end of the bunch $\Delta E_{max}$
with the cooperative action of CSR and Coulomb forces on the bunch end
\begin{equation}
\Delta E_{max}=LF_C(\sigma)+L_m *\hat\varepsilon '(1)
\label{coolcond}
\end{equation}
The cooling is possible, because compressed bunch has energy deviation. Notice, 
that the longitudinal CSR force heats, really, the bunch, because it is negative, 
but Coulomb force cools because it is positive and his action is negative to the 
compression direction. The bunch ends in the bend are moved to the bunch 
center, which is in fact the compression but the Coulomb force is directed in 
the opposite direction and his operation decreases the energy of the bunch 
ends, i.e. energy deviation. This more applicable to compressing of bunches 
of higher energy 
(
$\gamma  \geq 100$
) in the chicane, where the main part of the compressing way is in the free 
space and CSR is absent. In case of lower energy bunches 
(
$\gamma  < 100$
) the heating influence of the CSR is negligible in comparison to Coulomb and 
the compressing in any kind of alfa-magnet give us cooling of the bunch energy 
spread on approximately an order. The uncorrelated energy spread do not cooled 
by such simple cooling method. The cooling on one order of magnitude give 
us a possibility to increase the Madelung energy 
$\Gamma $
of the bunch by two orders of magnitude and the condition of liquid bunch \cite{1} is 
satisfies, if bunch has initially 
$\Gamma  > 0.01$
. This condition is possible to satisfy in various linear accelerator bunches. 
So, the liquid state of the bunch, where the bunch is partially ordered and can 
radiate coherent as laser \cite{1}, is possible to attaine at various linacs.

\section{ CONCLUSIONS}

We have analyzed the bunch Coulomb force and cooperative synchrotron 
radiation in the bending magnet and found that it essentially influences the 
microbunch dynamics. First of all, the longitudinal force redistributes energy 
losses along the bunch, so, that head particles are somewhat accelerated by 
the field radiated by tail particles. The energy losses originate from derivative 
of the linear charge density, that is characteristic feature of the effect. This 
acceleration give a possibility to decrease the energy deviation of the 
compressed bunch, i.e.. to cool the bunch. Under certain conditions the full 
energy deviation can be equals to zero after compression except uncorrelated 
energy spread. Aside of the energy losses the transverse focusing in the 
self space charge forces is possible.

\end{document}